\documentclass[11pt]{article}
\pdfoutput=1
\usepackage{latexsym}
\usepackage{amsmath,amssymb,amsfonts}
\usepackage{graphicx}
\usepackage{latexsym,color}
\usepackage{cite}

\title{\Large\textbf{Problems of interaction of a
supersonic gas mixture with a wall solved by the projection method
applied to the full Boltzmann equation}}

\author{\large\textbf{A. A. Raines}\\
St. Petersburg State University,\\ Universitetski pr. 28, St. Petersburg, 198504 Russia
\\ E-mail: raines@mail.ru}

\date{}

\begin{document}
\maketitle

\section{Introduction}

Our approach to problems of rarefied gas dynamics is based on
direct solution of the complete kinetic Boltzmann equation. The
main difficulty of solving the Boltzmann equation is the
calculation of the multi-dimensional collision integral. In this
paper, we apply our generalization of the conservative discrete
ordinate method of Tcheremissine \cite{Tcher} (which was
originally developed for a single gas) for binary gas mixtures in
the case of cylindrical symmetry (Raines A. A.) \cite{Raines}. For
the evaluation of collision integrals we use the projection method
which ensures the strict conservation of mass, momentum and
energy. The conservativeness of the method is achieved by a
special projection of the integrand values, calculated at non-node
points, to nodes of the momentum grid that are closest to them.

Using this method, we now solve the problem of interaction
of a two-component supersonic jet with a normally posed wall. We
discover the effects of the inflow of gas on a cool wall
with the mirror and diffuse reflection laws and the appearance of the wall
Knudsen layer. The results are compared with the
gas-dynamical solution and with the papers dealing with a similar
problem for a single gas \cite{Arist,Rykov}. We obtain a
good agreement with all those results.

\section{Description of the method}

The system of the Boltzmann equations in  the momentum space for
two gas components consisting of hard sphere molecules with the
masses $m_i$ and diameters $d_i$ reads
\begin{equation}
\frac{\partial f_i}{\partial t} +
\frac{\vec{p}}{m_i}\,\frac{\partial f_i}{\partial \vec{x}} = - L_i
+ G_i,\qquad i = 1,2,
 \label{1}
\end{equation}
where $f_i$ is the distribution function which depends on the
vector of momentum $\vec{p}$, the vector of configuration space
$\vec{x}$ and the time $t$.

Direct and reverse collision integrals have the form
\begin{equation}\label{2}
  L_i =
  \sum_{j=1}^2\int\limits_{-\infty}^{+\infty}\int\limits_0^{2\pi}\int\limits_0^{\pi}f_i f_{j^\ast}
  \frac{1}{2}\left(\frac{d_i+d_j}{2}\right)^2q_{ji}\sin{\theta}d
  \theta d\varphi d\vec{p}_\ast
\end{equation}
\begin{equation}\label{3}
  G_i =
  \sum_{j=1}^2\int\limits_{-\infty}^{+\infty}\int\limits_0^{2\pi}\int\limits_0^{\pi}f'_i f'_{j^\ast}
  \frac{1}{2}\left(\frac{d_i+d_j}{2}\right)^2q_{ji}\sin{\theta}d
  \theta d\varphi d\vec{p}_\ast
\end{equation}
\[ q_{ij} = |(\vec{g}_{ji}\cdot\vec{n})|.\]
In the kinetic momentum space we have the following relations between
the vectors of momentum before and after the collision
\begin{eqnarray}
\vec{p}' = \vec{p} + \frac{2m_i m_j}{(m_i+m_j)}\,
(\vec{g_{ji}}\cdot \vec{n})\vec{n} \nonumber
\\ \vec{p}'_\ast = \vec{p}_\ast - \frac{2m_i m_j}{(m_i+m_j)}\,
(\vec{g_{ji}}\cdot \vec{n})\vec{n}  \label{4}
\end{eqnarray}
\[\vec{g_{ji}} = \frac{\vec{p}_\ast}{m_j} - \frac{\vec{p}}{m_i} \]
\[\vec{n} = \vec{n}(\cos\theta,\sin\theta\cos\varphi,\sin\theta\sin\varphi) .\]
Here $\vec{n}$ is a unit vector directed along the interaction
line of molecules, $\vec{g}_{ji}$ is their relative velocity,
$\theta$ and $\varphi$ are collision angles.

We introduce in the limited domain of the configuration space a
fixed grid. We impose limits on the momentum variables in
(\ref{1}) - (\ref{3}) by introducing a domain $\Omega$ of volume
$V$. In $\Omega$ we construct a discrete grid containing $N_0$
equidistant points $\vec{p}_\beta$ with the step $h$ which results in the discretization
of the distribution functions and collision integrals as follows
\begin{equation}\label{5}
  f_i(\vec{p}^\ast) = \frac{V}{N_0}\sum\limits_{\beta =
  1}^{N_0}f_{i\beta}\delta(\vec{p}^\ast - \vec{p}_\beta)
\end{equation}
\begin{equation}\label{6}
  L_i(\vec{p}^\ast) = \frac{V}{N_0}\sum\limits_{\beta =
  1}^{N_0}L_{i\beta}\delta(\vec{p}^\ast - \vec{p}_\beta)
\end{equation}
\begin{equation}\label{7}
  G_i(\vec{p}^\ast) = \frac{V}{N_0}\sum\limits_{\beta =
  1}^{N_0}G_{i\beta}\delta(\vec{p}^\ast - \vec{p}_\beta)
\end{equation}
The Boltzmann equation in a discrete form becomes
\begin{equation}\label{8}
  \frac{\partial f_{i\beta}}{\partial t} +
  \frac{\vec{p}_\beta}{m_i} \frac{\partial f_{i\beta}}{\partial
  \vec{x}} = - L_{i\beta} + G_{i\beta}
\end{equation}
Equation (\ref{8}) is solved by the splitting procedure. On each
interval $\Delta t$ we split the process into the two stages,
free-molecular flow and collisional relaxation described by the following equations
\[\frac{\partial f_{i\beta}}{\partial t} +
  \frac{\vec{p}_\beta}{m_i} \frac{\partial f_{i\beta}}{\partial
  \vec{x}} = 0 \]
\[\frac{\partial f_{i\beta}}{\partial t} = - L_{i\beta} + G_{i\beta} \]
Let us consider the integral operator
\[\Omega_i(\Phi) = \sum\limits_{j=1}^2 \int\limits_{-\infty}^{+\infty}
\int\limits_{-\infty}^{+\infty}\int\limits_0^{2\pi}\int\limits_0^{\pi}\Phi
f_if_{j^\ast}\frac{1}{2}\left(\frac{d_i+d_j}{2}\right)^2
q_{ji}\sin\theta d\theta d\varphi d\vec{p}_\ast d\vec{p}
\]
Taking for $\Phi$ a three-dimensional $\delta$-function, we reduce
the collison integrals to the form
\begin{equation}\label{9}
L_i(\vec{p}^{\,\ast}) = \frac{1}{2}
\Omega_i\big(\delta(\vec{p}^{\,\ast} - \vec{p}) +
\delta(\vec{p}^{\,\ast} - \vec{p}_\ast)\big)
\end{equation}
\begin{equation}\label{10}
G_i(\vec{p}^{\,\ast}) = \frac{1}{2}
\Omega_i\big(\delta(\vec{p}^{\,\ast} - \vec{p}^{\,\prime}) +
\delta(\vec{p}^{\,\ast} - \vec{p}^{\,\prime}_\ast)\big)
\end{equation}
Transforming the integral operator to cylindrical coordinates and
taking into account the relations $\vec{p} =
\vec{p}(p,\rho,\gamma)$, $d\vec{p} = \rho dp d\rho d\gamma$ we obtain
\[\Omega_i(\Phi) = \sum\limits_{j=1}^2\, \int\limits_{\Omega\times\Omega}
\int\limits_0^{2\pi}\int\limits_0^{\pi}\Phi
f_if_{j^\ast}\frac{1}{2}\left(\frac{d_i+d_j}{2}\right)^2
q_{ji}\chi\sin\theta d\theta d\varphi d\vec{p}_\ast d\vec{p}
\]
Introduce the uniform integration grid: $p_{\alpha\nu}$,
$\rho_{\alpha\nu}$, $p_{\beta\nu}$, $\rho_{\beta\nu}$,
$\gamma_\nu$, $\gamma_{\ast\nu}$, $\theta_\nu$, $\varphi_\nu$ with
$N_\nu$ nodes. The multiple integral is calculated as the $8$-fold
sum over all the nodes while the distribution functions do not
depend on $\gamma_\nu$, $\gamma_{\ast\nu}$:
\begin{equation}\label{11}
  \tilde{L}_i(\vec{p}^{\,\ast}) =
  A\sum\limits_{\nu=1}^{N_\nu}\sum\limits_{j=1}^2
  J_\nu^{ij}\big(\delta(\vec{p}^{\,\ast} - \vec{p}_{\alpha_{\nu}})
  + \delta(\vec{p}^{\,\ast} - \vec{p}_{\beta_{\nu}})\big)
\end{equation}
\begin{equation}\label{12}
  \tilde{G}_i(\vec{p}^{\,\ast}) =
  A\sum\limits_{\nu=1}^{N_\nu}\sum\limits_{j=1}^2
  J_\nu^{ij}\big(\delta(\vec{p}^{\,\ast} - \vec{p}^{\,\prime}_{\alpha_{\nu}})
  + \delta(\vec{p}^{\,\ast} - \vec{p}^{\,\prime}_{\beta_{\nu}})\big)
\end{equation}
where
\[J_\nu^{ij} = f_{i\alpha_\nu}f_{j\beta_\nu}\frac{1}{2}\left(\frac{d_{i\alpha_\nu} +
d_{j\beta_\nu}}{2}^2 \right)q_{j\beta_\nu i\alpha_\nu}
sin{\theta_\nu}\rho_{\alpha_{\nu}}\rho_{\beta_{\nu}},\quad A =
V^2\pi^2/N_\nu \]
Combining (\ref{6}) and (\ref{11}) we obtain
\begin{equation}\label{13}
\tilde{L}_{i\beta} =
B\sum\limits_{\nu=1}^{N_\nu}\sum\limits_{j=1}^2 \left(
J_\nu^{ij\prime} + J_\nu^{ij\prime\prime}\right),\quad B =
V\pi^2/(N_\nu/N_0)
\end{equation}
We replace the expressions in parentheses in (\ref{12}) by using the relations
\begin{equation}\label{14}
  \delta(\vec{p}^{\,\ast} - \vec{p}^{\,\prime}_{\alpha_{\nu}})
+ \delta(\vec{p}^{\,\ast} - \vec{p}^{\,\prime}_{\beta_{\nu}}) =
\sum\limits_{\vec{s}} r_{\vec{s}}\big(\delta(\vec{p}^{\,\ast} -
\vec{p}_{\lambda_{\nu}+\vec{s}}) + \delta(\vec{p}^{\,\ast} -
\vec{p}_{\mu_{\nu}+\tilde{\vec{s}}})\big)
\end{equation}
The coefficients $r_{\vec{s}}$ can be found from the conditions of
conservation of the density, kinetic momentum and energy in the
decomposition (\ref{14}) for a pair of cells including their
vertices. For economy of computations it would be preferable to
have a decomposition with a minimum number of terms, so that
expression (\ref{14}) becomes
\begin{eqnarray}
& &  \delta(\vec{p}^{\,\ast} - \vec{p}^{\,\prime}_{\alpha_{\nu}})
+ \delta(\vec{p}^{\,\ast} - \vec{p}^{\,\prime}_{\beta_{\nu}}) = (1
- r_{\vec{s}^\ast})\big(\delta(\vec{p}^{\,\ast} -
\vec{p}_{\lambda_{\nu}}) + \delta(\vec{p}^{\,\ast} -
\vec{p}_{\mu_{\nu}})\big) \nonumber
\\ & & + r_{\vec{s}^\ast}\big(\delta(\vec{p}^{\,\ast} -
\vec{p}_{\lambda_{\nu}+\vec{s}^\ast}) + \delta(\vec{p}^{\,\ast} -
\vec{p}_{\mu_{\nu}+\tilde{{\vec{s}}^\ast}})\big)
 \label{15}
\end{eqnarray}
If we use (\ref{15}) in (\ref{12}) and combine with (\ref{7}),
then we obtain
\begin{equation}\label{16}
\tilde{G}_{i\beta} =
B\sum\limits_{\nu=1}^{N_\nu}\sum\limits_{j=1}^2 \left[\left(
J_\nu^{ij\prime} + J_\nu^{ij\prime\prime}\right)(1 - r_\nu) +
\left( J_\nu^{ij\ast} + J_\nu^{ij\ast\ast}\right)r_\nu \right]
\end{equation}
The expressions (\ref{13}) and (\ref{16}) define the conservative
discrete ordinate method if the coefficients have been already
found.

\section{Formulation of the problem}

Consider the problem of the inflow of a binary gas mixture upon a
wall with mirror and diffusive laws of reflection of the gas from
the wall. The half-space $x>0$ is being filled with the gas moving
with velocity $U$ (or momentum $P$) in the negative direction of
the $x$-axis. The density and temperature of the gas are equal to
$n_0$ and $T_0$, respectively. At the instant $t=0$ we set an
immovable wall at the point $x=0$. At $t>0$ the interaction of the gas
with the wall starts. The system of Boltzmann equations in
momentum space for this problem for the two gas components consisting
of hard sphere molecules with masses $m_i$ and diameters $d_i$ has
the form (\ref{1}).

\begin{figure}
\includegraphics[scale=0.5]{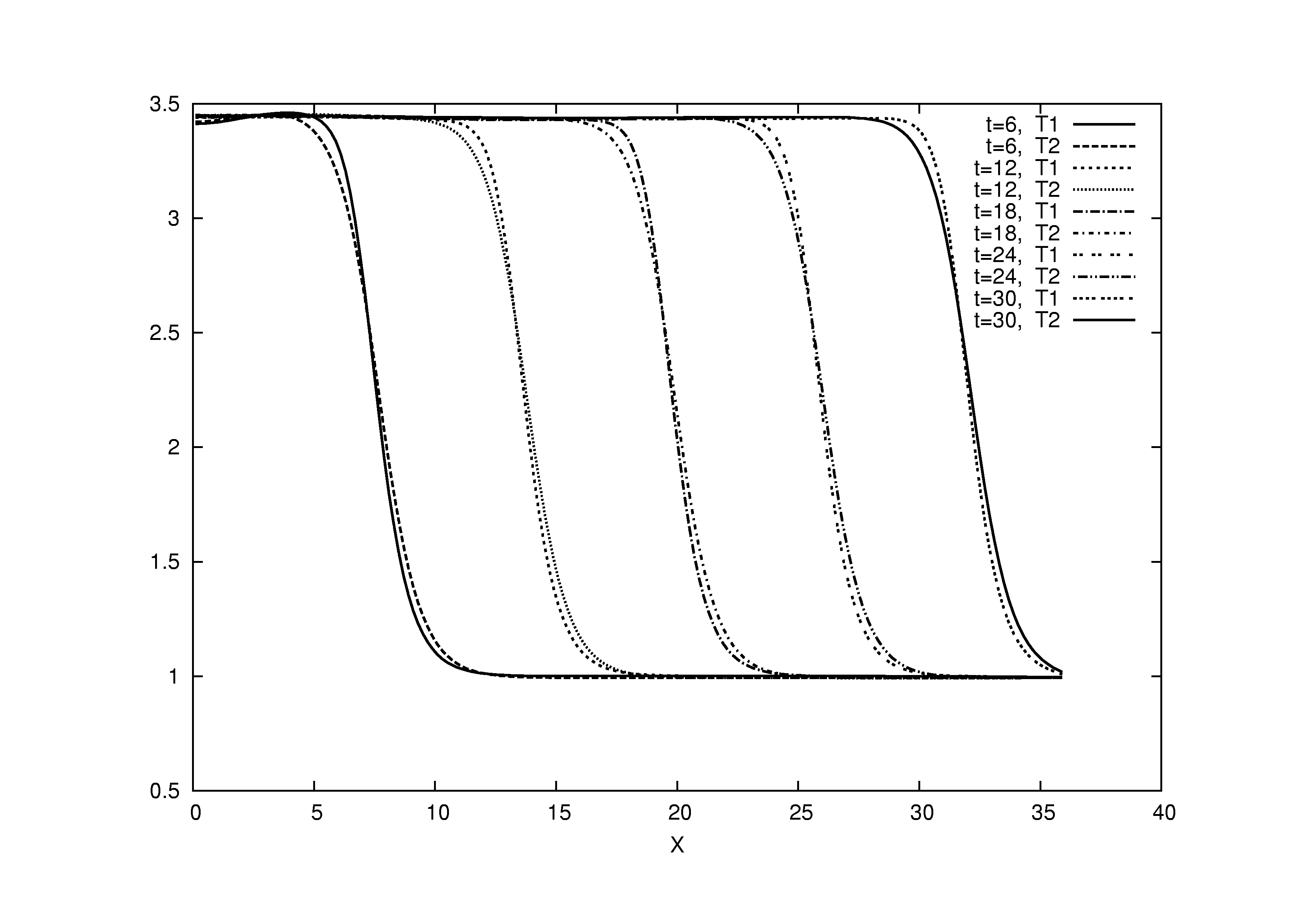}
\caption{Mirror reflection. Profiles of temperatures of the mixture components
at different times.\label{fig1}}
\end{figure}

\begin{figure}
\includegraphics[scale=0.5]{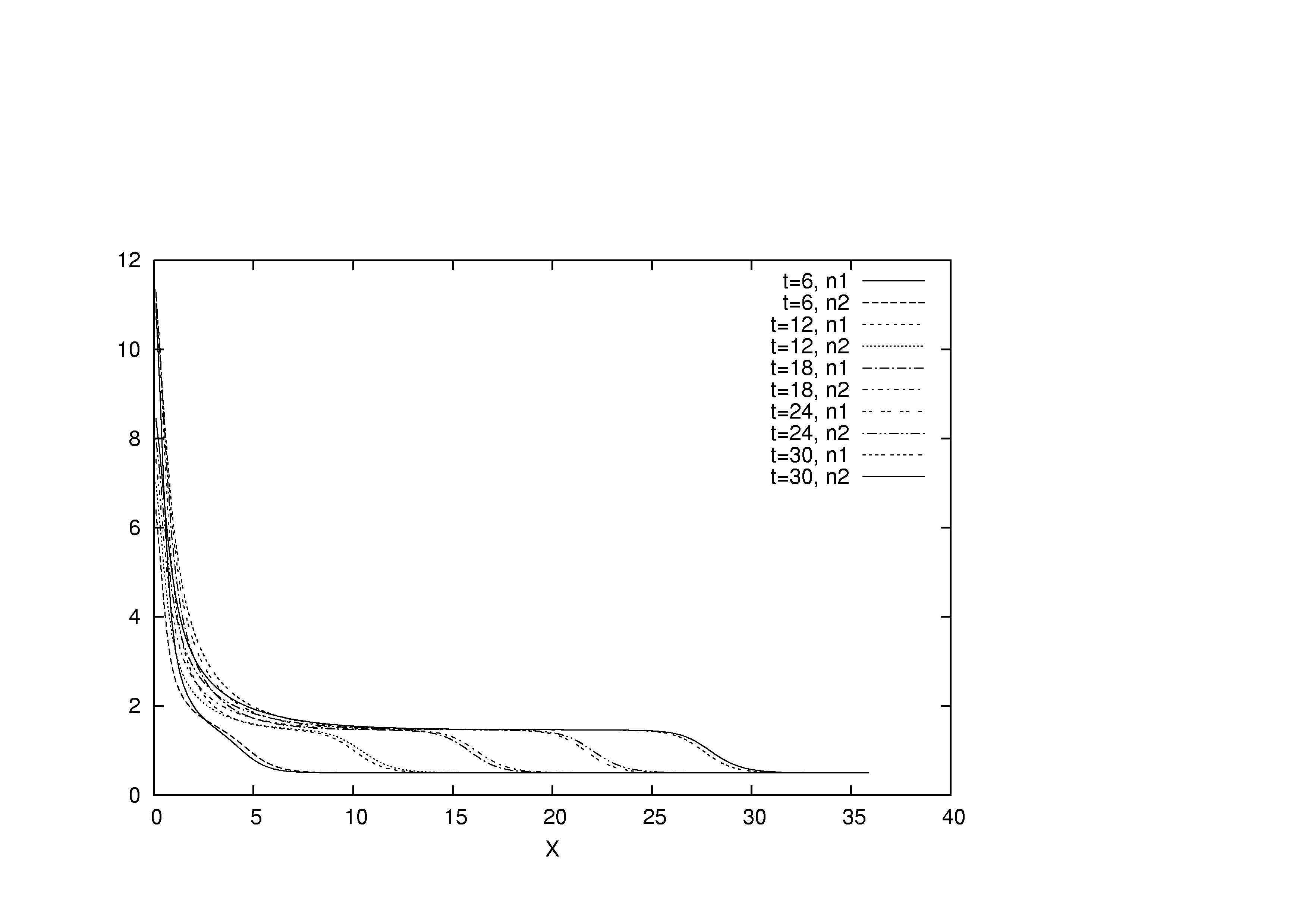}
\caption{Diffuse reflection. Profiles of numerical densities
 of the mixture components at different times. Temperature of the wall is equal to 0.5. 
 \label{fig2}}
\end{figure}

\begin{figure}
\includegraphics[scale=0.5]{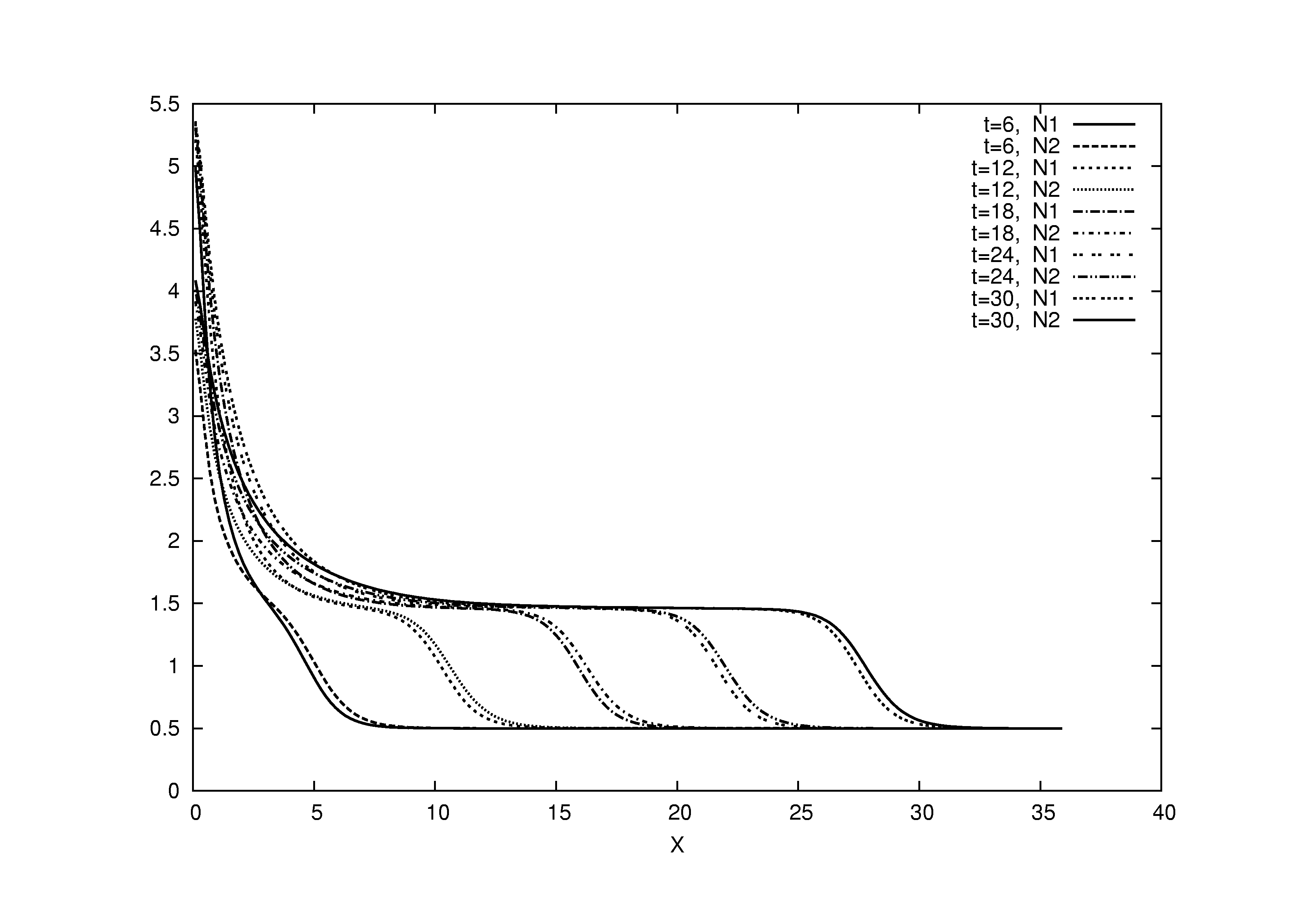}
\caption{Diffuse reflection. Profiles of numerical densities
 of the mixture components at different times. Temperature of the wall is equal to 1.
 \label{fig3}}
\end{figure}

\begin{figure}
\includegraphics[scale=0.5]{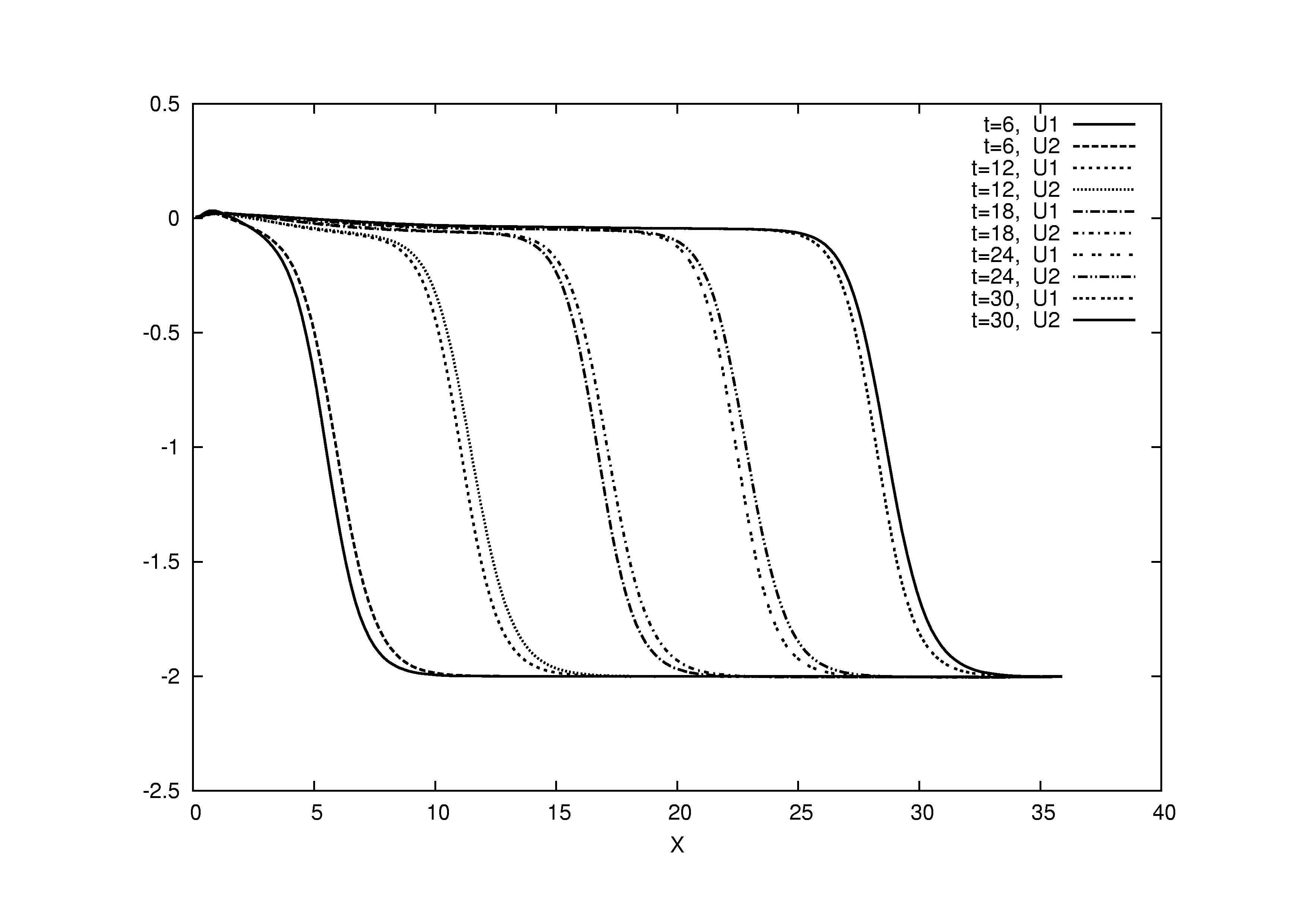}
\caption{Diffuse reflection. Profiles of velocities of the mixture components
 at different times. Temperature of the wall is equal to 1.\label{fig4}}
\end{figure}

\begin{figure}
\includegraphics[scale=0.5]{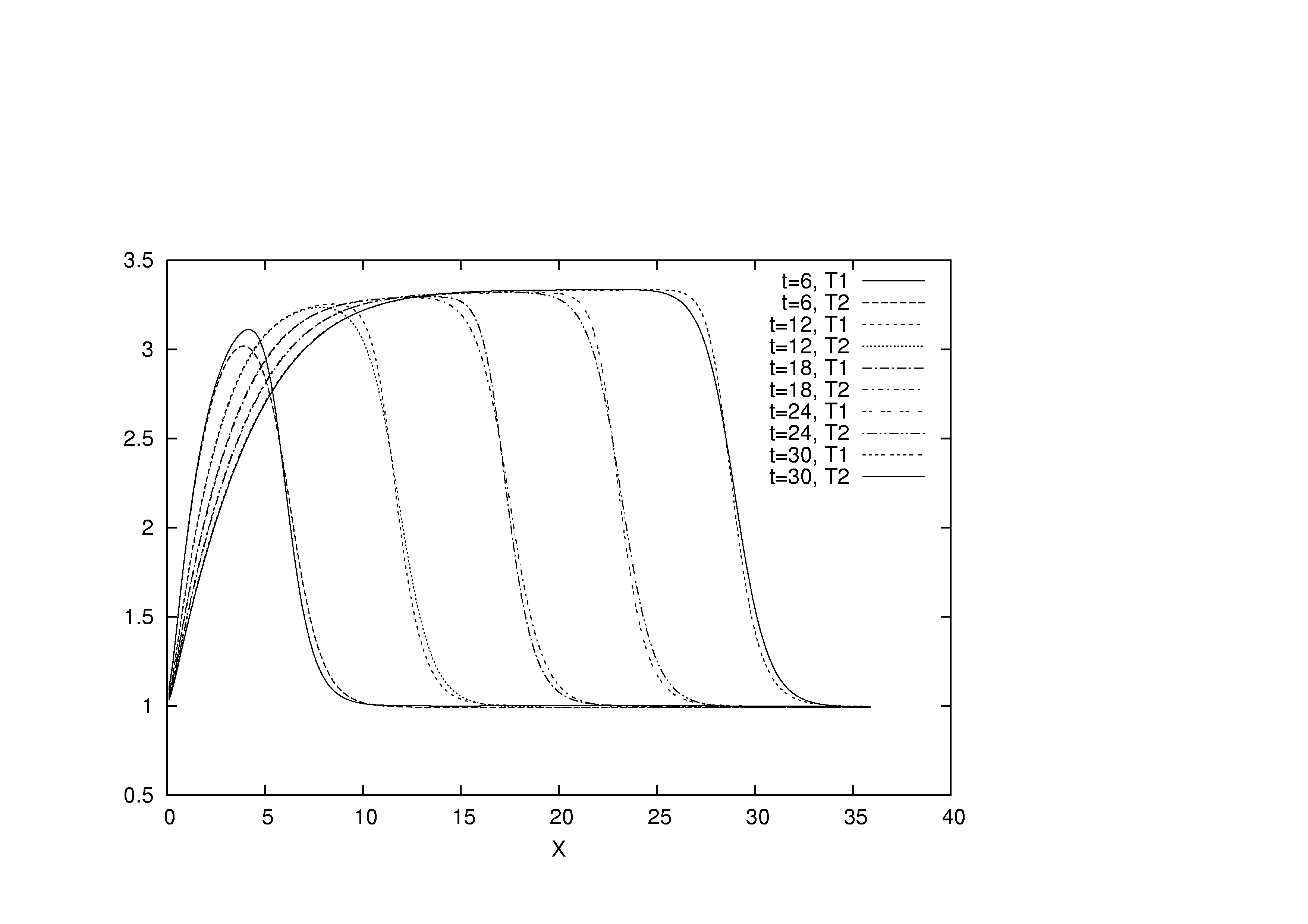}
\caption{Diffuse reflection. Profiles of temperatures of the mixture components
at different times. Temperature of the wall is equal to 1.\label{fig5}}
\end{figure}

\begin{figure}
\includegraphics[scale=0.5]{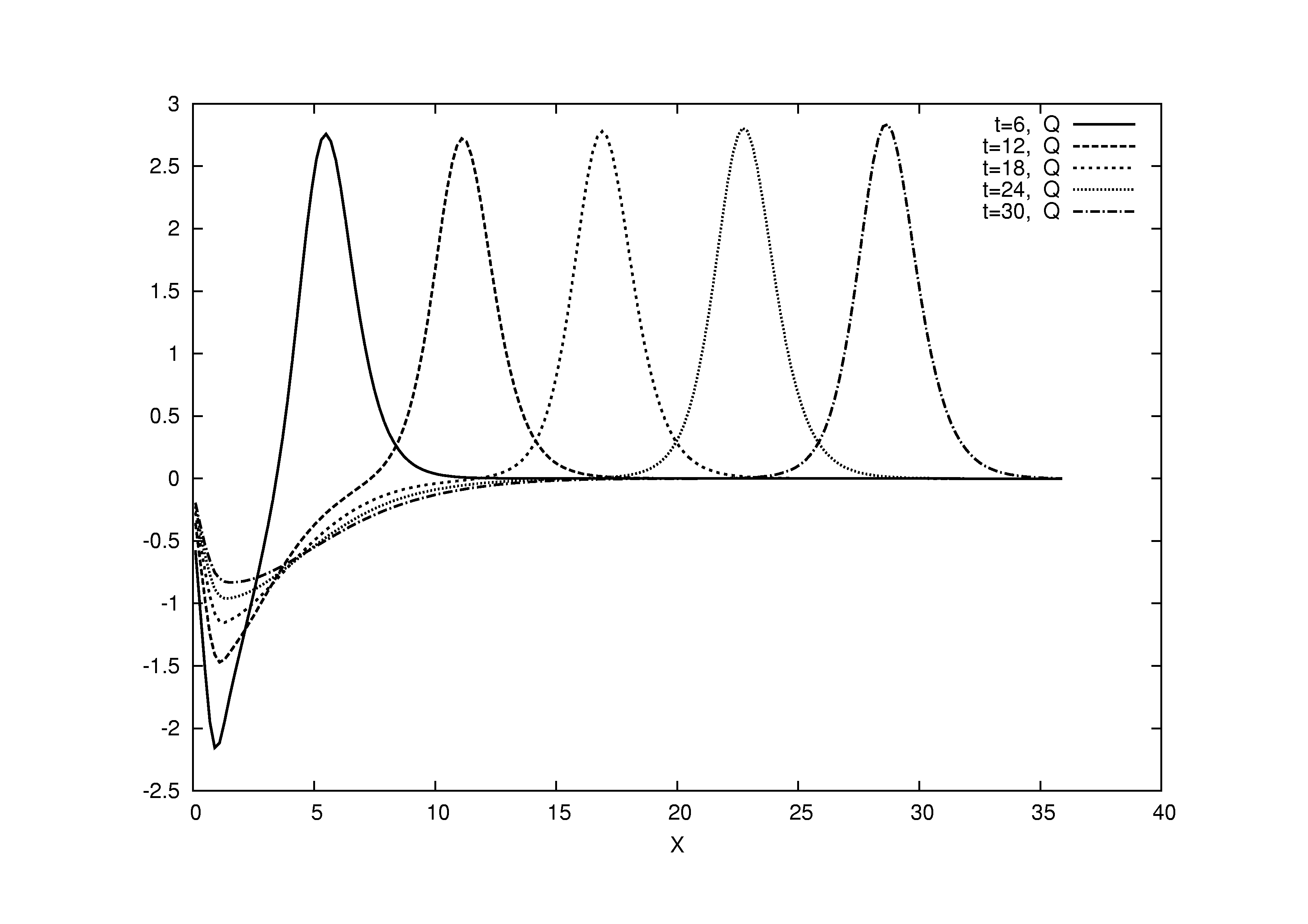}
\caption{Diffuse reflection. Profiles of the heat flow of the gas mixture 
at different times. Temperature of the wall is equal to 0.5.\label{fig6}}
\end{figure}

 A solution of the problem depends on four variables,
so that the distribution functions have the form $f_i =
f_i(t,x,p,\rho)$. As an initial condition for the distribution
functions we take Maxwell functions with parameters of the
unperturbed flow
\begin{equation}\label{maxwell}
  f_i(t=0,x,p,\rho) = n_{i0} \left(\frac{1}{2\pi
  kT_0m_i}\right)^{3/2} \exp{\left(-\frac{(p-m_iP/m)^2 +
  \rho^2}{2km_iT_0}\right)},
\end{equation}
where $n_{i0}$ is the number density of $ith$ component of the
mixture, $n_0 = n_{10} + n_{20}$, $m$ is mass of the mixture. For
the calculations, an infinite domain of the variation of $x$ is
replaced by the segment $[0,L]$ and at $x=L$ we set up conditions
(\ref{maxwell}) for molecules flying into the domain, which yields
a boundary condition on the free boundary surface
\begin{eqnarray}
 & & f_i\!\left(t=0,x=L,p<0,\rho\right) = n_{i0} \left(\frac{1}{2\pi
  kT_0m_i}\right)^{3/2} \times \nonumber
  \\ & & \qquad \qquad \qquad
  \exp{\left(-\frac{(p-m_iP/m)^2 +  \rho^2}{2km_iT_0}\right)}.
  \label{inflow}
\end{eqnarray}
On the left boundary of domain $[0,L]$ we set up conditions for
$p\geqslant 0$. In the case of mirror reflection from the wall,
the boundary conditions have the form
\begin{equation}\label{left}
  f_{im}\!\left(t=0,x,p\geqslant 0,\rho\right) = f_i(t,x,-p,\rho).
\end{equation}
For diffusive reflection with the full accommodation the
distribution functions of reflected molecules are assumed to be
maxwellian with the temperature equal to the wall temperature:
\begin{equation}\label{reflect}
  f_{iw}(t,x=0,p\geqslant 0,\rho) = n_{iw} \left(\frac{1}{2\pi
  kT_w m_i}\right)^{3/2} \exp{\left(-\frac{p^2 +
  \rho^2}{2km_iT_w}\right)}.
\end{equation}
Here $T_w$ is the wall temperature, $n_{iw}$ is the density of
reflected molecules which is found from the condition of
non-percolation:
$$
n_{iw} = -
\frac{2\sqrt{\pi}}{\sqrt{2km_iT_w}}\iiint\limits_{p<0}f_ip\,d\vec{p}.
$$
Thus, the problem of the inflow of a rarefied gas on a wall is
posed for system (\ref{1}) with initial conditions
(\ref{maxwell}), boundary conditions (\ref{inflow}) on the free
boundary surface and boundary conditions (\ref{left}),
(\ref{reflect}) on a solid wall. Graphical representation of solution results
can be seen in figures \ref{fig1}--\ref{fig6}.

\section{Dimensionless quantities} 
Here we introduce the dimensionless quantities corresponding to
all physical variables which are involved in our solution method.
\begin{eqnarray*}
&c_0 = \sqrt{\frac{2kT_0}{m_1}} , \qquad p_0 = \sqrt{2kT_0 m_1}&
\\ &\lambda = 1/(\sqrt{2}\pi d_1^2 n_0),\qquad \tau = \lambda/c_0&
\\ &\vec{p}^{\,\prime} = \vec{p}/p_0,\quad \vec{u}^{\,\prime}/c_0,\quad
t'=t/\tau,\quad X' = x/\lambda,\quad f_i/(n_0p_0^{-3}) = f_i'&
\\ &n' = n/n_0,\quad T' = T/T_0,\quad q' = q/(n_0p_0^3m_1^{-2})&
\\ & P' = - \frac{3\sqrt{5}}{4\sqrt{6}}\frac{(M^2-1)}{M}\sqrt{M_x}, \quad
 U' =  - \frac{3\sqrt{5}}{4\sqrt{6}}\frac{(M^2-1)}{M}\frac{1}{\sqrt{M_x}}.&
 \\ &M = Mach,\qquad M_x = m_{mix}/m_1. &
\end{eqnarray*}

\section{Parameters of calculations}

All calculations were performed at the laptop computer Sony VAIO,
processor Intel(R) Core(TM) 2CPU, 1.66GHz + 1.66GHz, 1.00GB of
RAM.

Time for an iteration step is 2.4 sec., time for the whole problem
until $t=30\tau$ is 120min ($\tau$ is the mean free time).

Numerical parameters of calculations:
\begin{eqnarray*}
&M = 2.8773,\quad U = - 2,\quad m_2/m_1 = 0.5,&
\\ &n_{2_0}/n_{1_0} = 0.5/0.5,\quad d_2/d_1 = 1, Kn = 1.&
\end{eqnarray*}
The grids used for calculations:
 \newline $Np\rho = 1250 = (25 +
25)25$ - the number of nodes of the momentum grid with the step $h
= 0.2$
 \newline $NX = 180$ - the number of nodes of the $x$-grid with
 the step $h_x = 0.2$. \newline
 $NOD = 66000$ integration nodes, $DT = 0.01$ is the time step.

\section{Comparison of numerical results with the gas-dynamical
solution}
According to gas dynamics equations, a domain is created
with constant values of macroscopic parameters behind the reflected shock wave.
For parameters specified in section 5 this gas-dynamical solution is the following
\[n_{gd} = 2.936,\quad n_{1gd} = n_{2gd} = 1.468,\quad T_{gd} = 3.4396 \]
Here $n_{gd}, n_{1gd}, n_{2gd}$ are numerical densities of the mixture, first component,
and second component, respectively, while $T_{gd}$ is the temperature of the mixture.

Our results obtained by the direct numerical solution of the Boltzmann equation are described below.

1. Mirror reflection.\newline
 When $T_{wall} = T_{gas} = 1$, there is the zone of
 constant values for density and temperature
 \begin{eqnarray*}
 &n = 2.93598,\quad n_1 = 1.48607,\quad n_2 = 1.451&
 \\ &T = 3.44089,\quad T_1 = 3.44135,\quad T_2 = 3.4375. &
 \end{eqnarray*}
2. Diffusive reflection.\newline
 a) When $T_{wall} = 0.5$, $T_{gas} = 1$ and $t > 18\tau$, $x > 12.5\lambda$, we have the region of
 constant values for density and temperature
 \[n = 2.9348,\qquad T = 3.3435. \]
 b) When $T_{wall} = T_{gas} = 1$ and $t > 18\tau$, $x > 12.5\lambda$ we have the
 region of constant values for density and temperature
 \[n= 2.9313,\qquad T = 3.3135. \]
 We observe a good agreement of our results with the gas-dynamical solution.

 \section{Comparison of our results with the results obtained by other methods}

\begin{figure}
\includegraphics[scale=0.5]{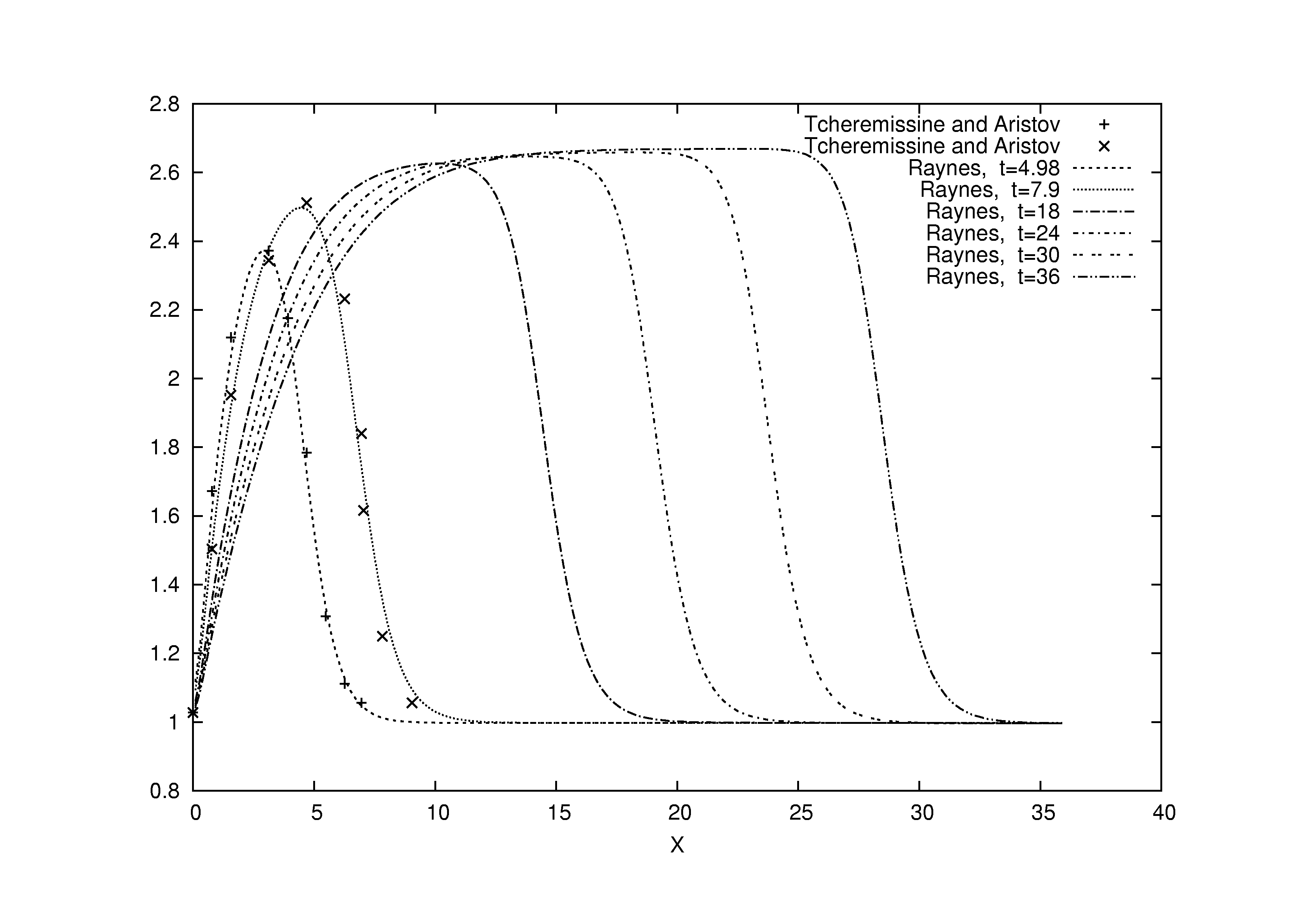}
\caption{Comparison of results of Raines (Raynes)
with the paper \cite{Arist} for temperature profiles for a one-component gas.\label{fig7}}
\end{figure}

\begin{figure}
\includegraphics[scale=0.5]{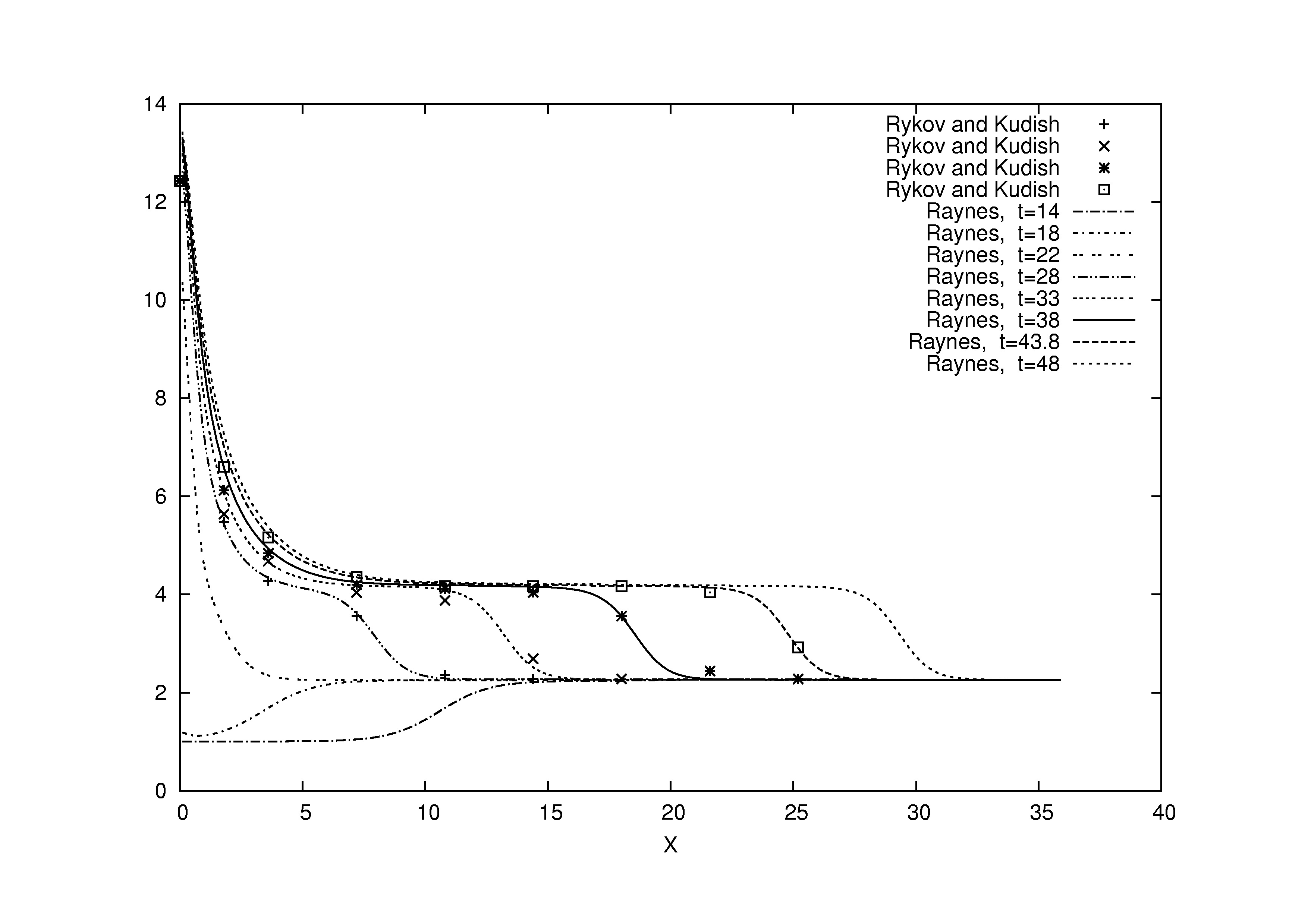}
\caption{The problem of reflection of a shock wave from a wall. 
Comparison of results of Raines (Raynes)
with the paper \cite{Rykov} for density profiles for a one-component gas.\label{fig8}}
\end{figure}

\begin{figure}
\includegraphics[scale=0.5]{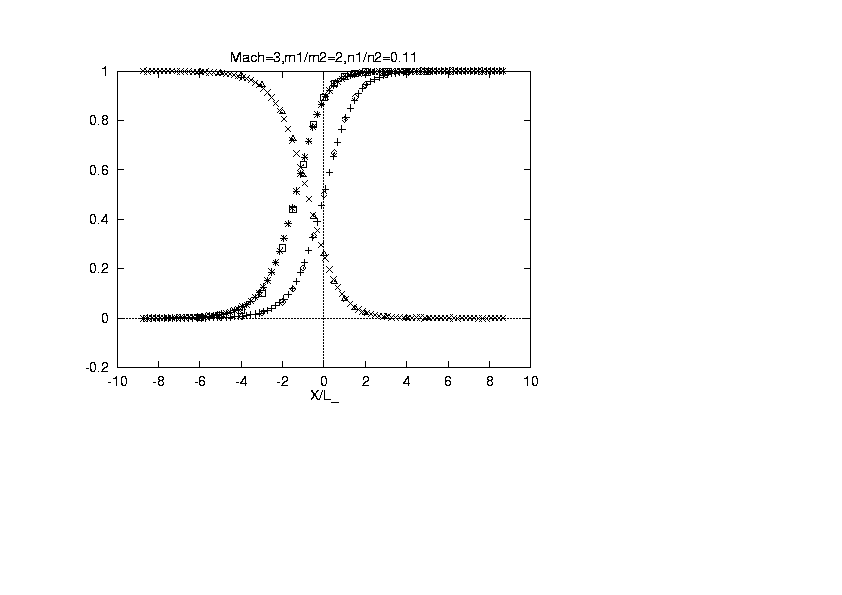}
\caption{The problem of a shock wave in a binary gas mixture.
Comparison with results of \cite{Kosuge}. Profiles of density, velocity and temperature 
for the mixture.\label{fig9}}
\end{figure}

 Among other publications studying this problem, we should mention two earlier works that restrict themselves
 to a one-component gas \cite{Arist,Rykov}. In \cite{Arist}, a conservative difference scheme is applied
 for solving the Boltzmann kinetic equation by the splitting method, with a good agreement of results 
 shown in Figure \ref{fig7}. In \cite{Rykov} a scheme of the second order of accuracy with respect to $\Delta t$ 
 is applied on the basis  of the model Krook equation. For the sake of comparison with \cite{Rykov} we have solved
 the problem of reflection of a shock wave from a wall with a reasonable agreement of our result with \cite{Rykov}
 shown in Figure \ref{fig8}. Initial test of the projection method was made on the problem of a shock wave 
 in a binary gas mixture where a comparison of results was made with the work \cite{Kosuge} for a wide range of parameters.
 An example of such comparison is shown in Figure \ref{fig9}.


\end{document}